\documentclass[twocolumn]{openjournal}
\usepackage{makecell}
\usepackage{amsmath}

\defcitealias{Planck2016}{Planck Collaboration et al. 2016}
\usepackage{hyperref}
\hypersetup{
    colorlinks = true,
    linkcolor = blue,
    citecolor = blue,
}
\begin{document}

\title{%Constraints on Early Supermassive Black Hole Growth from Halo-Scale Environments
Early Supermassive Black Holes and Little Red Dots Require Free-Fall Growth %from Halo-Scale Environments
}

\author[0000-0002-6038-5016]{Junehyoung Jeon}
\affiliation{Department of Astronomy, University of Texas, Austin, TX 78712, USA}
\affiliation{Cosmic Frontier Center, The University of Texas, Austin, TX 78712, USA}
\author[0000-0003-0212-2979]{Volker Bromm}
\author[0000-0002-9604-343X]{Michael Boylan-Kolchin}
\affiliation{Department of Astronomy, University of Texas, Austin, TX 78712, USA}
\affiliation{Cosmic Frontier Center, The University of Texas, Austin, TX 78712, USA}
\affiliation{Weinberg Institute for Theoretical Physics, University of Texas, Austin, TX 78712, USA}
\author[0000-0002-8984-0465]{Julian B. Mu{\~n}oz}
\affiliation{Department of Astronomy, University of Texas, Austin, TX 78712, USA}
\affiliation{Cosmic Frontier Center, The University of Texas, Austin, TX 78712, USA}
\affiliation{Weinberg Institute for Theoretical Physics, University of Texas, Austin, TX 78712, USA}

\email{junehyoungjeon@utexas.edu}

\begin{abstract}
Supermassive black holes/active galactic nuclei (SMBHs/AGN), forming only a few hundred million years after the Big Bang as observed with the James Webb Space Telescope (JWST), challenge theoretical understanding. How could they grow so massive $(M_{\rm BH} > 10^6 {\rm \,M}_\odot)$ so quickly after initial seeding? Is this rapid growth related to the numerous and enigmatic Little Red Dots (LRDs), compact sources with AGN-like characteristics, discovered by JWST? To address these mysteries, we consider the first-order constraint on SMBH growth: enough baryonic material has to reach the vicinity of the SMBH seed, located near the bottom of the gravitational potential well of the host dark matter halo. We specifically examine cold-mode accretion, where gas from the cosmic environment flows into the virialized halo in cold streams without being shock-heated, efficiently reaching the center on a free-fall timescale. We find that cold mode accretion is \textbf{\textit{necessary}} to supply material for the SMBHs to reach the observed masses, whereas for shock-heated gas inflow the required amount could only be supplied by implausibly rare halos. Moreover, cold-mode inflow in rare $(\sim1$ Gpc$^{-3}$) halos matches the mass and number of the massive quasars, and halos able to support super-Eddington accretion for massive SMBHs ($\sim10^7$ M$_\odot$) match LRD number densities. The decreasing LRD abundance at lower redshifts may then reflect the termination of cold-mode accretion in the growing host halos. The populations of massive SMBHs and LRDs at early times may thus arise naturally from cosmological structure formation, based on the abundance of halos capable of supplying sufficient material through cold accretion.
\end{abstract}

\keywords{Early universe — Supermassive black holes — Active galactic nuclei — Theoretical models}

\section{Introduction} \label{sec:intro}

How do supermassive black holes (SMBHs) emerge in the early Universe \citep[e.g.,][]{Smith2019_2, Woods2019,Inayoshi2020}? Massive quasars $(>10^9$ M$_\odot)$ and active galactic nuclei (AGN; $\sim10^6-10^8$ M$_\odot$) have been observed in the first billion years of cosmic history \citep{Wu2015}, now reaching back to $\lesssim500$~Myr after the Big Bang \citep[e.g.,][]{Bogdan2023,Maiolino2023,Napolitano2024,Taylor2025,Chavez2025}. Moreover, the James Webb Space Telescope (JWST) has observed a larger abundance of high-$z$ AGN than previously expected \citep[e.g.,][]{Kocevski2023,Onoue2023,Furtak2023,Greene2023,Taylor2024}. For such massive objects to emerge so early, the SMBH seeds must have formed even earlier, combined with rapid growth over extended periods \citep[e.g.,][]{Jeon2023,Jeon2024,Chon2026}.

%The formation and evolution of supermassive Black Holes (SMBHs) in the early Universe have posed a longstanding challenge in astrophysics \citep{Smith2019_2,Woods2019,Inayoshi2020}. Massive quasars $(>10^9$ M$_\odot)$ have been observed in the first billion years of cosmic history \citep{Wu2015}, and with the James Webb Space Telescope (JWST) an abundant population of massive active galactic nuclei (AGN) have been discovered \citep[e.g.,][]{Kocevski2023,Onoue2023,Furtak2023,Greene2023,Taylor2024}. The current frontier extends to $z\gtrsim10$, to less than $\sim500$ Myr after the Big Bang \citep[e.g.,][]{Bogdan2023,Maiolino2023,Napolitano2024}. To reach such masses so early, the SMBH seeds must have formed even earlier, combined with rapid growth over extended periods \citep[e.g.,][]{Jeon2023,Jeon2024,Chon2026}. 

Multiple theoretical scenarios have been proposed to solve the SMBH mass assembly challenge. One invokes super-Eddington accretion onto black hole (BH) seeds, accreting more efficiently than the fiducial Eddington limit \citep{Volonteri2021,Jeon2025}. Another is the heavy seed scenario, starting from a more massive BH $(\sim10^4-10^6$ M$_\odot$), compared to a stellar remnant, to mitigate the growth required to reach the observed AGN masses. Different heavy seed channels have been proposed: The direct-collapse black hole (DCBH) pathway produces a massive BH seed (\citealt{Becerra2018b}) under rare conditions that can suppress low-temperature gas cooling, so that a massive cloud collapses without vigorous fragmentation into stars with a supermassive star (SMS) as an intermediate, short-lived stage \citep[e.g.,][]{Bromm2003,Begelman2006,Lodato2006,Johnson2013,Haemmerl2018}. To prevent fragmentation, low metallicity \citep[$Z\lesssim 10^{-3}\ \rm Z_\odot$;][]{Chon2024} and/or sustained high accretion/infall rates of $\sim1$ M$_\odot$ yr$^{-1}$ may be required \citep{Regan2020,Wise2019,Nandal2026}. Runaway collisions of (proto-)stars/BHs in extremely dense stellar clusters \citep{Reinoso2023,Gaete2024}, or primordial BHs (PBHs) formed through collapsing overdensities soon after the Big Bang \citep[][]{Dayal2024,Zhang2025,Guo2026} could also produce heavy BH seeds. Lastly, hierarchical BH mergers could produce massive SMBHs at early times \citep[e.g.,][]{Bhowmick2024}. However, all three scenarios -- super-Eddington accretion, heavy seeds, or mergers -- still require efficient gas accretion onto the seed BH to reach the observed AGN masses that are orders of magnitude larger \citep{Jeon2024,Zhou2026}.

% 10^-3 to 10^-2
%$, heavy seed scenarios have been proposed. Among them is the direct-collapse black hole (DCBH) pathway, based on the runaway collapse of a massive, extremely metal-poor \citep[$Z\lesssim 10^{-3}\ \rm Z_\odot$;][]{Chon2024} gas cloud, and involving a supermassive star (SMS) as an intermediate, short-lived stage \citep[e.g.,][]{Bromm2003,Begelman2006,Lodato2006}. This channel produces a more massive BH seed ($\sim10^4-10^6$ M$_\odot$; \citealt{Becerra2018b}), and requires rare conditions that suppress low-temperature gas cooling mechanisms, allowing the cloud to collapse without fragmenting into a large number of ordinary/low-mass stars \citep[e.g.,][]{Johnson2013,Wise2019,Haemmerl2018,Haemmerle2020}. Sustained high accretion/infall rates of $\sim1$ M$_\odot$ yr$^{-1}$ could establish the conditions to form an SMS that soon collapses to a DCBH \citep{Regan2020,Wise2019,Nandal2026}. 
%Other scenarios for heavy seed BH scenarios exist, such as runaway collisions of (proto-)stars/BHs in extremely dense stellar clusters \citep{Reinoso2023,Gaete2024} or Primordial BHs (PBHs) formed through collapsing overdensities soon after the Big Bang \citep[][]{Dayal2024,Zhang2025}. We focus on the DCBH pathway in this work.

The JWST discovery of a previously unidentified population of objects known as ``Little Red Dots" (LRDs) presents an additional, and possibly related, early-Universe puzzle. These common and compact objects show multiple permitted broad lines in their spectra, indicative of gas kinematics around a central AGN \citep[e.g.,][]{Matthee2023,Kokorev2024_lrd}. However, many lack features commonly found in local AGN such as X-ray emission \citep{Akins2024,Kocevski2023} and variability \citep{Liu2026}. An emerging model for the nature of LRDs proposes that they host an accreting central SMBH surrounded by dense, dust-free gas \citep[e.g.,][]{Inayoshi2025_bl,Naidu2025}. The dense gas produces stellar-like signatures in the observed spectrum, producing strong breaks and exponential line profiles \citep{deGraff2025,Naidu2025,Kokorev2025}. Still, the required hydrogen densities are extreme $(10^8-10^{10}$ cm$^{-3})$ and large amounts of gas will be needed to support such dense configurations. Other proposed models to explain the lack of X-ray and variability include super-Eddington accreting AGN \citep{Madau2026,LiuH2025,Secunda2026,Pacucci2024} or SMSs \citep{Chisholm2026,Begelman2025}, all requiring massive gas inflow.

%Lastly, JWST has discovered a new population of objects, named the Little Red Dots (LRDs). These common and compact objects show multiple unambiguous permitted broad lines, expected to originate from strong gas outflows or rotation caused by an AGN \citep{Kocevski2023,Matthee2023,Kokorev2024_lrd,Akins2024}. However, they lack features commonly found in local AGN including weak X-ray emission and variability. An emerging model for the nature of LRDs is that of an accreting central SMBH surrounded by dense, dust-free gas \citep{Inayoshi2025_bl}. The extremely dense gas imprints signatures in the observed spectrum similar to that of a stellar population, producing strong breaks and exponential line profiles observed in LRD spectra \citep{deGraff2025,Naidu2025,Kokorev2025}. However, the required hydrogen densities are extreme $(10^8-10^{10}$ cm$^{-3})$ and large amounts of gas will be needed to support such dense configurations. Another proposed LRD model to explain the lack of X-ray and variability is that of super-Eddington accreting AGN \citep{Madau2026,LiuH2025,Secunda2026,Pacucci2024}, which will also require efficient gas inflow.

The challenges and proposed solutions mentioned above, i.e., the growth of BH seeds to SMBHs, the formation of heavy seeds, and the feasibility of LRD super-Eddington/dense gas configurations, all require to first order that large amounts of baryonic material are efficiently transported to the centers of dark matter halos. Previous studies have investigated gas inflow inside halos at larger galaxy-scales in the context of fueling star and SMBH formation \citep[e.g.,][]{Keres2005,Keres2009,Birnboim2007,Dekel2006,Dekel2009,Yoshida2003,Fernandez2014,Wise2019,Regan2020,Latif2022}. With the advent of JWST and the discovery of abundant high-redshift galaxies and AGN, the question of efficiently forming dense gas regions to encourage star formation and SMBH growth has been revisited \citep{MBK2023,Boylan2025,Dekel2025,Dekel2023,Prole2024,Prole2024_2}. Following such work, we present an analytic model of baryonic accretion inside dark matter halos at small scales near halo centers in the early Universe, to test how likely or numerous environments are with sufficiently large baryonic infall to support the emergence of massive SMBHs at early times, and whether the observed SMBH population can naturally be explained by the matter supply from cosmological structure formation. At early times, with higher halo gas density and inflow, BH fueling may be more efficient, enabling SMBH formation.

% what is different from starburst calculation? Is there different behaviro at high redshifts for smbhs?

%The abovementioned questions, accretion and growth of SMBH seeds, formation of DCBHs, and feasibility of LRD super-Eddington/dense gas configurations, all require to first order that large amount of baryonic material be transported efficiently to the centers of dark matter halos where these SMBHs are expected to reside. Previous works have investigated gas inflow inside halos in context of star formation \citep[e.g.][]{Keres2005,Keres2009,Birnboim2007,Dekel2006,Dekel2009}. With the advent of JWST and the discovery of abundant high-redshift galaxies and AGN, the question of efficiently forming dense gas regions to encourage star formation and SMBH growth have been revisited \citep{MBK2023,Boylan2025,Dekel2025,Dekel2023}. Following such work, we present an analytic model of baryonic accretion inside dark matter halos in the early Universe, to test how likely or numerous are environments with large enough baryonic infall to support the evolution of massive SMBHs as being observed at early times, and whether the observed SMBH population can naturally be explained by the material supply from cosmological structure formation. 

In this work, we adapt \textit{Planck} cosmological parameters \citepalias{Planck2016}: $\Omega_{\rm m} = 0.315$, $\Omega_{\rm b} = 0.048$, $\sigma_8 = 0.829$, $n_{\rm s} = 0.966$, and $h = 0.6774$. The cosmic baryon fraction is therefore $f_{\rm b}\equiv\Omega_{\rm b}/\Omega_{\rm m}=0.15$. We introduce the halo accretion model in Section~\ref{sec:model}, describe the resulting abundances of efficiently accreting systems in Section~\ref{sec:masses}, discuss the comparison with observations in Section~\ref{sec:discussion}, and summarize our work in Section~\ref{sec:conclusions}.

%\begin{figure*}[!htb]
%\gridline{
%\fig{halo_accretion.png}{0.5\textwidth}{}
%\fig{halo_accretion_10.png}{0.5\textwidth}{}
%}
%\caption{The internal halo accretion rate using the Bondi-Holye and free-fall models at $z=20$ and 10 for various halo masses. The virial radii of the halo are in physical units. While the Bondi-Hoyle model is larger than the optimal free-fall rate at large radii $(\gtrsim10^{-2}R_{vir})$, the Bondi-Hoyle halo accretion model may not be relevant near the halo outskirts. We plot the total profile for completeness and to compare against the regime where high halo density inhibits feedback \citep{Boylan2025} and the halo accretion rate \citep{Fakhouri2010}. The free-fall rate is comparable to the halo growth rate calibrated from simulations \citep{Fakhouri2010} near the virial radius.}
%    \label{fig:accmodel}
%\end{figure*}
% need a formula for T_0?
\begin{figure*}
        \centering
    \includegraphics[width=0.8\textwidth]{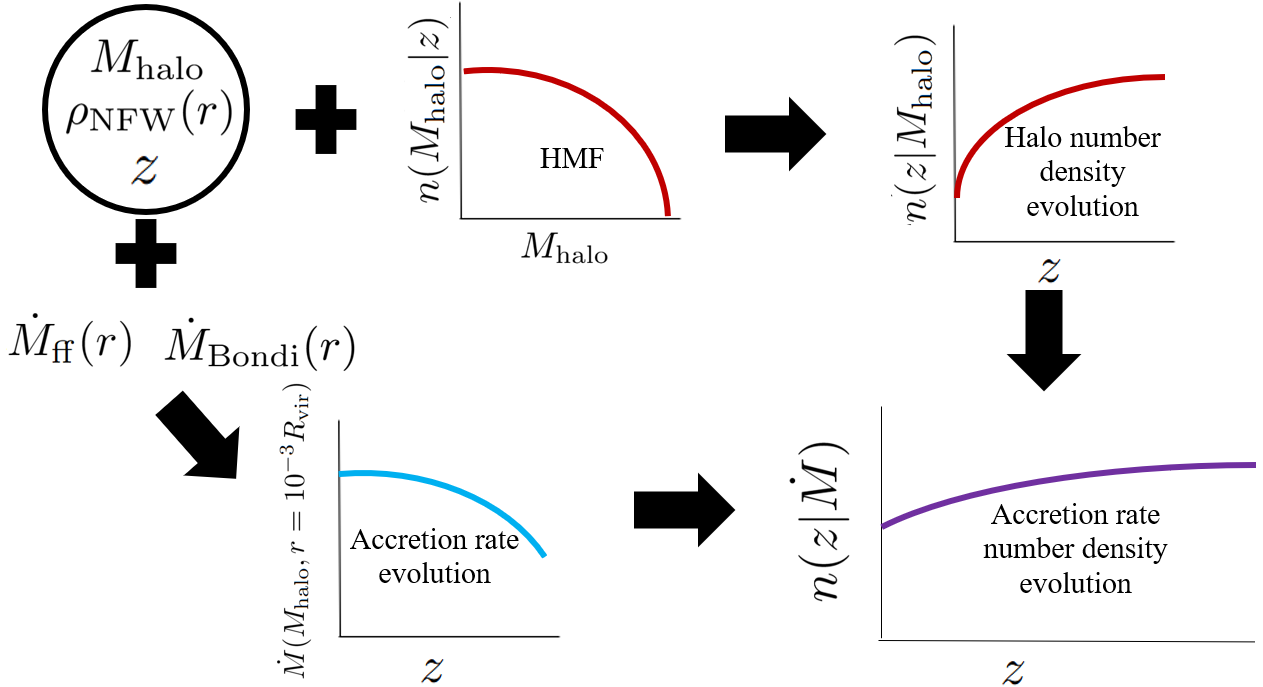}
    \caption{Conceptual framework for predicting the number density of systems that can support a given accretion rate across redshifts. For a specific halo mass and redshift, the infall rate at a given radius (here $10^{-3}R_{\rm vir}$) can be determined based on Eq.~\ref{eq:ff} or \ref{eq:bondi}. Together with the halo mass function (HMF), the volume density of such systems can be inferred. Combining these ingredients, we arrive at the redshift-evolution of accretion rate volume densities, as discussed in detail in the main text.}
    \label{fig:outline}
\end{figure*}

\section{Halo Inflow Model} \label{sec:model}

When gas infalls into a virialized dark matter halo, it will generally be shock heated to the virial temperature of the halo, then eventually cool to be able to collapse \citep{White1978,Mo2010}. However, more efficient free-fall cold-mode accretion is also possible, where the gas is not shock heated but remains relatively cold $(\sim10^4$ K) during inflow through filaments and streams for less massive $(\lesssim10^{11}-10^{12}$ M$_\odot$) halos, where shock heating is not significant \citep{Dekel2009,Dekel2006,Birnboim2007,Keres2005}. To produce the observed high-redshift SMBHs, we thus consider two modes of accretion: the cold-mode free fall model, where we assume that the gas infalls at the ideal free fall rate, compared against the Bondi-Hoyle accretion model \citep{Bondi1944}, which is dependent on the halo temperature profile and assumes that the gas is initially heated to the halo virial temperature. 

\subsection{Cold Mode Free-Fall Accretion}

For cold mode accretion, we assume that it proceeds at free fall, representing the most efficient rate to transport matter \citep{Hobbs2012}:
\begin{equation}
    \dot{M}_{\rm ff}(r) = \frac{f_{\rm b}M_{\rm enc}(r)}{t_{\rm ff}(r)}\mbox{\ ,}
    \label{eq:ff}
\end{equation}
where $M_{\rm enc}(r)$ is the mass enclosed within radius $r$ and the free fall time is $t_{\rm ff}(r) = \sqrt{3\pi/[32G\rho(r)]}$, with $G$ being the gravitational constant and $\rho(r)$ the halo density profile. To determine $M_{\rm enc}(r)$, we adopt the Navarro-Frenk-White (NFW) dark matter halo density profile \citep{Navarro1996}.
\begin{equation}
    \rho_{\rm NFW}(r) =  \frac{\rho_0}{\frac{r}{R_{\rm s}}\left(1+\frac{r}{R_{\rm s}}\right)^2}\mbox{\ ,}
\end{equation}
where $\rho_0$ is the characteristic density, $R_{\rm s}$ the scale radius, defined as $R_{\rm s} = R_{\rm vir}/c_{\rm halo}$, with $c_{\rm halo}$ being the halo concentration, and $R_{\rm vir}$ the virial radius. For the halo concentration, we use the model of \citet{Ishiyama2021}.

%Fig.~\ref{fig:accmodel} compares the resulting free-fall against the mass accretion rate of dark matter halos calibrated from simulations \citep{Fakhouri2010}. 

Even though the halo density could allow for such efficient accretion, stellar feedback effects might disrupt it. In this context, it has been argued that the gravitational potential provided by virialized dark matter can cause high gravitational accelerations/surface density of the gas, capable of suppressing the effect of stellar feedback \citep{Boylan2025}. Relatively rare ($\sim10^{-4}$ Mpc$^{-3}$) and massive halos at high redshifts ($\gtrsim10^{10}$ M$_\odot$ at $z\sim10$) have been predicted to show such suppression. For such halos, the gas infall should not be strongly affected by feedback at early times, and our optimistic halo accretion estimates may be achievable. As a further consistency check, we find that our values of $\dot{M}_{\rm ff}$ match the accretion rates near the halo virial radius \citep{Fakhouri2010}.

%Fig.~\ref{fig:accmodel} compares the free-fall rate with the Bondi-Hoyle rate. We also mark the mass accretion rate of dark matter halos calibrated from simulations \citep{Fakhouri2010} and the dark matter dominated regime of high density and inefficient feedback to disrupt the dense gas in halos \citep{Boylan2025}. The halo accretion rates are comparable to the free-fall model accretion rates. An oddity is that the Bondi-Hoyle model is higher at large radii $(\gtrsim10^{-2}R_{vir})$ than the optimal free-fall rate. However, as stated earlier, we expect the Bondi-Hoyle model to apply more near the dense halo center and not the outskirts. Still, the most optimal $\dot{M}_{ff}$ matches halo accretion rate near the halo virial radius. Furthermore, at $z=20$, massive halos (except 10$^{8}$ M$_\odot$ case) have regions of high enough density so that stellar feedback will not inhibit gas collapse. Thus, our optimistic halo accretion estimates may be possible.

\subsection{Bondi-Hoyle Accretion}
As a comparison to the cold mode free-fall model, we consider the Bondi-Hoyle formalism to model the shock heated gas inflow. Whereas this formalism is designed for spherically symmetric accretion from a uniform gas distribution onto a point-like object, implying that it might not apply to the halo on large scales, it is adequate for the smaller scales relevant for gas accretion at the dense center of a halo \citep{Hobbs2012}. We define the Bondi-Hoyle accretion rate at halo radius $r$ as
\begin{equation}
    \dot{M}_{\rm Bondi}(r)=4\pi\frac{[GM_{\rm enc}(r)]^2\rho_{\rm NFW}(r) f_{\rm b}}{c_{\rm s}(r)^3}\mbox{\ ,}
    \label{eq:bondi}
\end{equation}
where the sound speed is 
\begin{equation}
    c_{\rm s} = \sqrt{\frac{\gamma k_{\rm B} T(r)}{\mu m_{\rm p}}}\mbox{\ .}
\end{equation}
Here, $\gamma$ is the polytropic index, $k_{\rm B}$ the Boltzmann constant, $T(r)$ the gas temperature profile, $\mu=1.22$ the mean molecular weight of the primordial gas, and $m_{\rm p}$ the proton mass.  

For the temperature profile, we adapt the model of \citet{Komatsu2001}: 
\begin{equation}
    T(x) = \frac{T_0}{\left[x\left(1+x\right)^2\right]^{\gamma-1}}\mbox{\ ,}
\end{equation}
where $x = r/R_{\rm s}$. $T_0$ is defined as 
\begin{equation}
    T_0 = T_{\rm vir}\eta_0 = \frac{G\mu m_{\rm p} M_{\rm vir} \eta_0}{3 R_{\rm vir} k_{\rm B}}\mbox{\ ,}
\end{equation}
where $M_{\rm vir}$ is the halo virial mass and 
\begin{equation}
    \eta_0 = 0.00676(c_{\rm halo}-6.5)^2 + 0.206(c_{\rm halo}-6.5)+2.48\mbox{\ .}
\end{equation}
We also adopt $\gamma=1.15+0.01(c_{\rm halo}-6.5)$ from \citet{Komatsu2001}.

%Fig.~\ref{fig:accmodel} shows the resulting Bondi-Hoyle accretion rate profiles. An oddity is that the Bondi-Hoyle model is higher at large radii $(\gtrsim10^{-2}R_{vir})$ than the optimal free-fall rate. However, as stated earlier, we expect the Bondi-Hoyle model to apply more near the dense halo center and not the outskirts. 

We note that in our model, we have not considered the effects of circular velocity/angular momentum, which the shock-heated gas might have during infall. Therefore, the shock-heated inflow rate could be much lower in reality. However, as shown below (Sec.~\ref{sec:masses}), even our optimistic Bondi-Hoyle model has difficulty reproducing observations.\footnote{On the other hand, at the smallest scales, the baryons may be more concentrated than dark matter, thus boosting the inflow rate. Such effects cannot be well determined by analytical models however, and future work will be needed.}

\begin{figure*}[!htb]
\gridline{
\fig{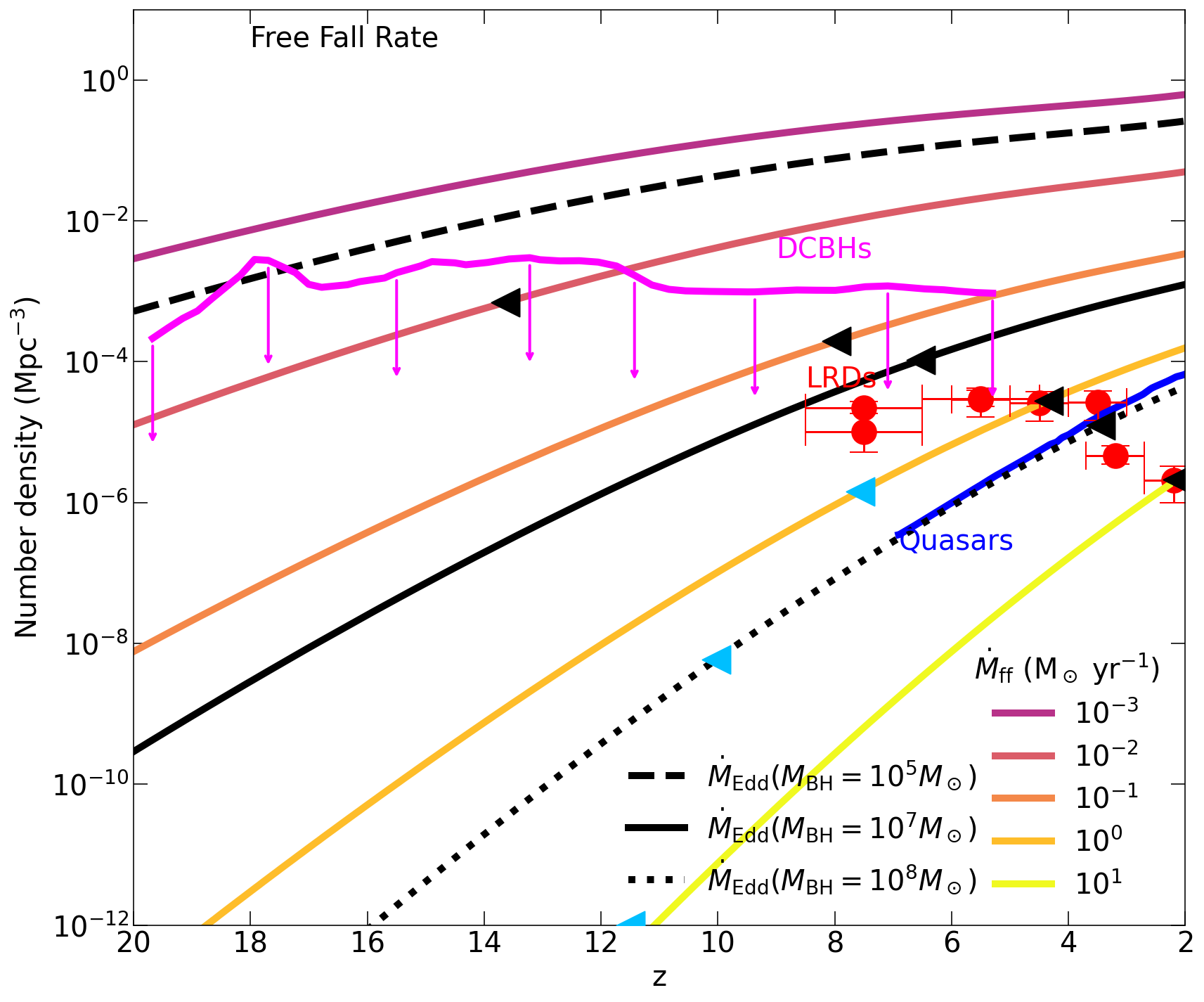}{0.5\textwidth}{}
\fig{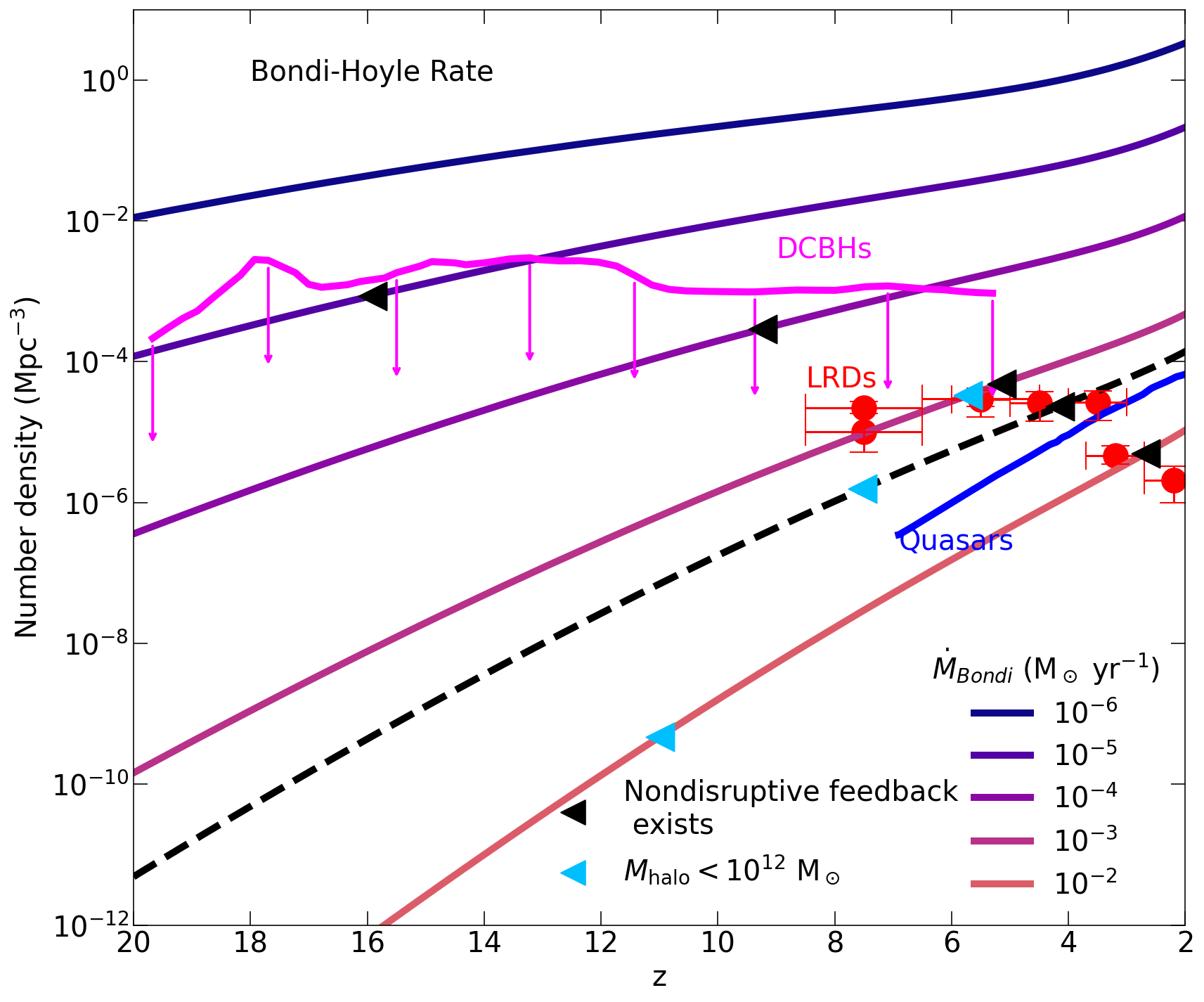}{0.5\textwidth}{}
}
\caption{Number density of systems with a given accretion rate (at $10^{-3}R_{\rm vir}$) 
%number density 
across redshift. {\it Left panel:} Case of free-fall accretion. {\it Right panel:} Situation for Bondi-Hoyle rate. For reference, we also show isocurves for Eddington accretion rates onto $10^5$, $10^7$, and $10^8$ M$_\odot$ SMBHs, assuming that free-fall/Bondi-Hoyle infall produces the given Eddington rate. In both panels, we further indicate the regime for non-disruptive feedback \citep{Boylan2025} with black arrows, and the fiducial limit for cold-mode accretion, corresponding to $M_{\rm halo}<10^{12}$ M$_\odot$ halos \citep{Keres2005,Dekel2006,Birnboim2007}, with blue arrows. We note that the accretion rate needed to form an SMS/DCBH ($\sim1$ M$_\odot$ yr$^{-1}$) is rare even for the optimal free fall conditions. We compare against the number densities inferred for high-redshift quasars \citep{Shen2020}, LRDs \citep{Kocevski2025,Kokorev2024_lrd,Zhuang2025,Ma2026}, and the predicted upper limit of Eddington-accreting DCBHs \citep{Jeon2025}. The quasar abundance matches the density for extreme accretion rates, as expected. The LRD number density is lower (corresponding to higher accretion rates) than $\dot{M}_{\rm ff}=\dot{M}_{\rm Edd}(M_{\rm BH}=10^7$ M$_\odot$), and more abundant (less accreting) than $\dot{M}_{\rm Bondi}=\dot{M}_{\rm Edd}(M_{\rm BH}=10^5$ M$_\odot$). Therefore, if LRDs are super-Eddington accreting AGN, near free-fall accretion must apply for most LRDs. The lack of LRDs at lower redshifts could then reflect the termination of cold-mode accretion, thus not being able to support super-Eddington accretion and/or dense gas envelopes as the LRD host halos grow in mass. The predicted DCBH number density matches the Eddington rate for the canonical DCBH seed mass, $10^5$ M$_\odot$, assuming free-fall accretion at high redshifts. The DCBH abundance stays roughly constant afterwards, implying that many of them will experience prolonged dormant phases. 
%Then, the number of efficiently evolving DCBHs likely stayed constant throughout cosmic history,%. This number at lower redshifts does not correspond to the number density of extreme accretion rates. Thus, 
%many likely remaining dormant or having dormant phases.
}
    \label{fig:accnumber}
\end{figure*}

\section{Cosmic History of SMBH Fuelling} \label{sec:masses}
Given the halo accretion models, we examine their contribution to SMBH growth across cosmic time and compare against observations. Specifically, we consider three related aspects: The number densities of systems that can support a given accretion rate, the total accreted masses for different number densities, and the resulting SMBH mass functions. In this analysis, we further explore two important regimes regarding the accretion model: the presence of halos with dark matter gravitational potential that can allow efficient gas collapse even under feedback \citep{Boylan2025}, and the fiducial limit of when free-fall (cold mode) accretion is possible, for halo masses $M_{\rm halo}\lesssim10^{12}$ M$_\odot$ \citep{Keres2005,Dekel2006,Birnboim2007}. The redshift range where both regimes overlap can be considered as the period for possible efficient gas inflow.

% lrd shuts off without cold mode accretion?

\subsection{Occurrence of Different Accretion Regimes}\label{sec:ratedis}
% more explanation?
We examine how many halos can support high accretion rates across cosmic time, employing the methodology outlined in Fig.~\ref{fig:outline}. We start with the  Sheth-Tormen halo mass function \citep[HMF;][]{Sheth1999}, as implemented in the Colossus package \citep{Diemer2018}. The HMF provides the number density of different mass halos at a given $z$, $n(M_{\rm halo} | z)$, and the number density of a given halo mass across $z$, $n(z| M_{\rm halo})$. The free-fall and Bondi-Hoyle accretion models of Eq.~\ref{eq:ff} and \ref{eq:bondi} give a mass inflow rate for a given halo mass and radius, $\dot{M}(M_{\rm halo},r)$. Both models also depend on the NFW profile and $M_{\rm enc}$ which evolve with $z$. Thus, given a halo mass (and a choice of halo radius at which to determine the inflow rate), its number density and inflow rate across redshifts can be computed. Combining them, we arrive at the cosmological number density of systems that can support a given accretion rate. 
%Using the Sheth-Tormen halo mass function \citep[HMF,][]{Sheth1999} implemented in the Colossus package \citep{Diemer2018}, 
We choose $r\simeq10^{-3}R_{\rm vir}$ to determine the mass accretion rate, as this corresponds to $\sim1$ pc for the halo masses typically encountered at $z\sim10-20$ ($M_{\rm halo}\gtrsim10^8$ M$_\odot$). 
%The number density is determined by matching the accretion rate at a given redshift and halo mass to the halo's number density based on the HMF. 
In Fig.~\ref{fig:accnumber}, we show the number density evolution for given infall rates, $\dot{M}(r=10^{-3}R_{\rm vir})$, across cosmic time. 
As reference cases, we also show the number densities of systems that could support accretion at the Eddington limit: 
\begin{equation}
\dot{M}_{\rm Edd} = 2.7\times10^{-3}\left(\frac{M_{\rm BH}}{10^5~\text{M}_\odot}\right)\left(\frac{\epsilon_r}{0.1}\right)^{-1}\rm~M_\odot~\text{yr}^{-1}\mbox{\ .}
    \label{edd}
\end{equation}
We consider select SMBH masses, $M_{\rm BH}=10^5,10^7,10^8$ M$_\odot$, assuming that free-fall (Eq.~\ref{eq:ff}) or Bondi-Hoyle (Eq.~\ref{eq:bondi}) infall produces the given $\dot{M}_{\rm Edd}$.

% dcbh plot all from model or just ones that can be observed?

We compare the predicted number densities to observations of LRDs \citep{Kocevski2025,Kokorev2024_lrd,Zhuang2025,Ma2026}, a model fit to high-redshift quasar observations \citep{Shen2020}, and to predictions for DCBHs accreting at the Eddington limit \citep{Jeon2025}. For rare/massive halos, the nondisruptive feedback regime persists even to $z\sim5$, thus providing some justification for our idealized treatment here, where gas infall is assumed to be largely unaffected by stellar feedback. 

Examining Fig.~\ref{fig:accnumber}, we find that under Bondi-Hoyle accretion, even the rarest and most massive halos with the lowest number densities cannot support efficient accretion. Thus, to fuel massive SMBH growth, cold-mode accretion will be necessary. When compared against free fall accretion, the quasar number density is close to the prediction for systems accreting at the Eddington rate onto a $10^8$ M$_\odot$ SMBH, in line with the extreme nature of quasars. However, the halos corresponding to $\dot{M}_{\rm Edd}(M_{\rm BH}=10^8$ M$_\odot$) exceed the cold-mode limit of $M_{\rm halo}\sim10^{12}$ M$_\odot$ at $z\sim10$, before the periods when these quasars are observed $(z\sim5-6)$. %This indicates that quasars could have gained their mass in earlier periods compared to the period we are observing them, are accreting at sub-Eddington rates, agreeing with existing high-redshift quasar observations \citep{Shen2019,Farina2022}. 

Regarding LRDs, at $z\gtrsim5$, under free-fall they are rarer than systems experiencing Eddington accretion rates for $10^7$ M$_\odot$ SMBHs, but more common than the prediction for $M_{\rm BH}\simeq 10^8$ M$_\odot$. This is in agreement with the idea that LRDs involve super-Eddington SMBHs, assuming that they harbor $M_{\rm BH}\lesssim10^8$ M$_\odot$, and that their host halos support free-fall accretion. We further see that the accretion rate matching LRD number densities ($\sim$1~M$_\odot$ yr$^{-1}$) corresponds to the $M_{\rm halo}\lesssim10^{12}$ M$_\odot$ regime near the period when the observed LRD number density is the highest, agreeing with LRD host halo masses from clustering observations \citep{Lin2025,Carranza2025}, and further roughly follows the limit of nondisruptive feedback towards lower $z$. The declining LRD number densities at lower redshifts may thus arise from the lack of gas inflow to support compact dense gas configurations/super-Eddington accretion.

Compared to predicted Eddington-accreting DCBH number densities, we find that the high-redshift DCBH abundance corresponds to the Eddington rate for the canonical DCBH seeding mass of $10^5$ M$_\odot$, and stays roughly constant. At lower redshifts, this number corresponds to a free-fall rate of $\sim10^{-1}$ M$_\odot$ yr$^{-1}$. Given such a modest rate, we infer that many DCBHs remain dormant after formation and do not grow to extreme masses. %to low accreting systems, even for the free-fall model. Then, to form DCBHs by accretion alone will be difficult should this upper limit hold. Other channels such as Lyman-Werner radiation destroying molecular hydrogen in preexisting massive gas clouds to prevent fragmentation and allowing gas collapse to form DCBHs may be necessary. In turn, most of the DCBHs may remain dormant, if their number density is as high as the predicted upper limit to match the low accretion rates predicted for the high number density systems. 

\begin{figure}
    \centering
    \includegraphics[width=\linewidth]{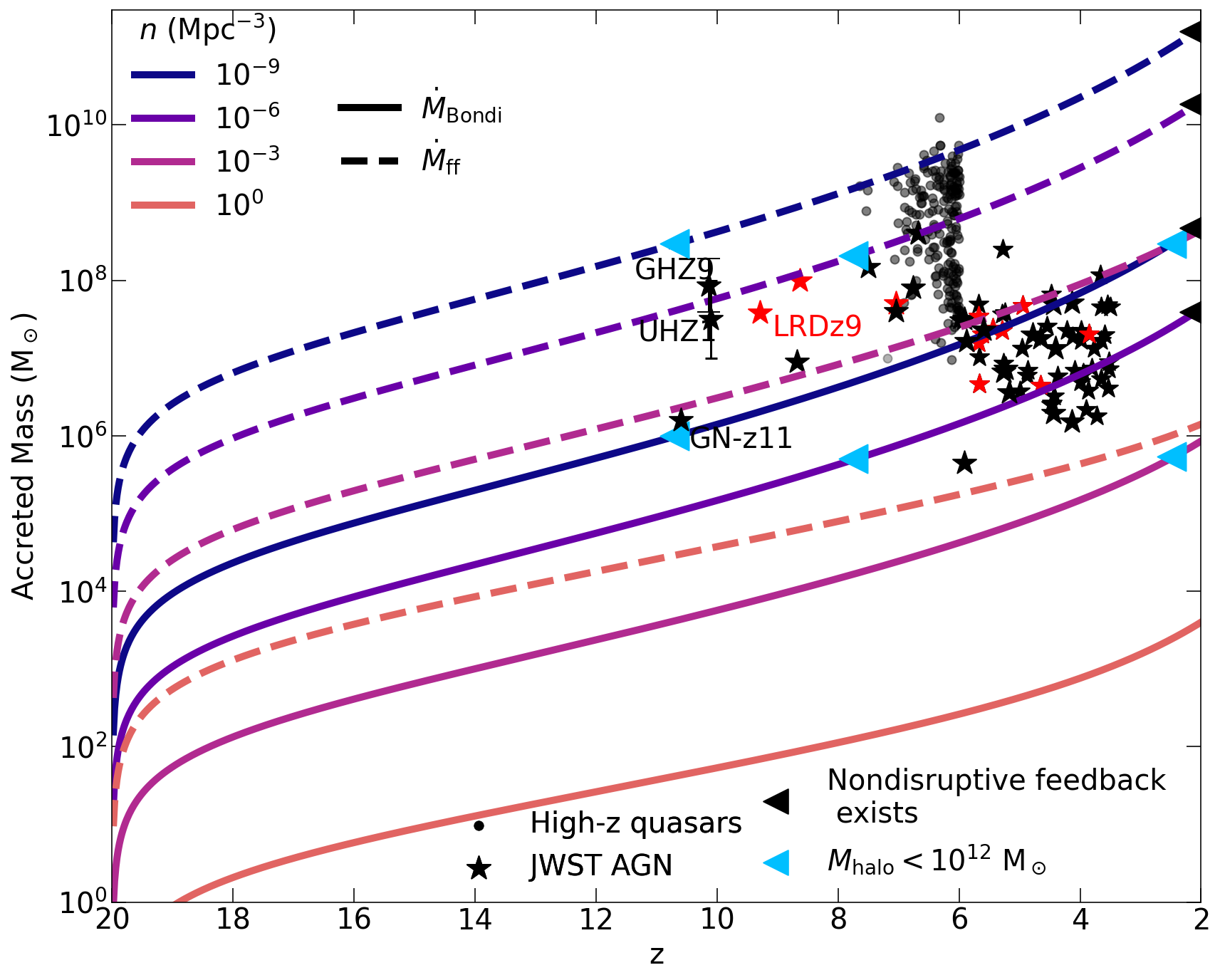}
    \caption{Cumulative mass growth across redshift assuming that halos evolve with constant number density. Using the Bondi-Hoyle or free fall accretion rate at $10^{-3}R_{\rm vir}$, we calculate the total mass accreted across redshift for a given halo number density. We compare the results against high-redshift quasar and AGN observations, including LRDs marked as red stars. The regimes for non-disruptive feedback and cold-mode accretion ($M_{\rm halo}<10^{12}$ M$_\odot$) are marked as in Fig.~\ref{fig:accnumber}. Both the observationally inferred quasar and JWST AGN masses could be reached by free-fall accretion, although for the most massive quasars, cold-mode accretion may not be possible anymore towards lower redshifts.}
    \label{fig:mvsz}
\end{figure}
% the masses here are upper limits. when does cold inflow regime stop?

\subsection{Cumulative SMBH Mass Growth} \label{sec:accmass}

To derive the total accreted mass across redshifts, we assume that halos evolve with constant number densities. Following halo masses corresponding to the same number densities in the HMF across redshifts, we determine the accreted mass up to each redshift, again using the accretion rate at $10^{-3}R_{\rm vir}$.  Fig.~\ref{fig:mvsz} shows the resulting mass evolution for different halo number densities. As we start at $z=20$, there may be additional mass that could have been accreted previously. However, as shown in the figure, soon after $z\sim20$ the mass grows exponentially, such that any mass accreted at $z>20$ is not significant.

We compare the accreted masses from our models with those inferred for high-redshift quasars and AGN observed by JWST \citep{Taylor2024,Taylor2025,Tripodi2024,Larson2023,Bogdan2023,Maiolino2023_2,Kokorev2023,Furtak2023,Juodbalis2024,Napolitano2024,Wang2010,Willott2017,Decarli2018,Izumi2018,Pensabene2020,Inayoshi2020,Fujimoto2022}. We find that to supply enough gas to grow the observed SMBHs, free-fall accretion is generally necessary. Under Bondi-Hoyle accretion, halos rarer than $\sim1$ Gpc$^{-3}$ are needed to produce the observed AGN, a much lower number density than even those inferred for LRDs. Under free fall accretion, the quasars match the rarest systems, where the most massive quasars can be produced by halos with number densities of $\sim1$ Gpc$^{-3}$. However, again, such rare systems reach the cold-mode accretion limit of $M_{\rm halo}\sim10^{12}$ M$_\odot$ before the accreted mass reaches the quasar range. This in fact matches the clustering measurements of the quasar host halos with masses $\gtrsim10^{12}$ M$_\odot$ \citep{Eilers2024}. Less massive AGN, observed by JWST, could be produced by more common halos under free-fall accretion. The JWST AGN, even the highest redshift ones at $z\sim10$, are less massive and can be assembled by free-fall accretion before the cold mode shuts off, agreeing with clustering measurements of their host halos being $\lesssim10^{12}$ M$_\odot$ \citep{Arita2025}.

% which halos can support free fall accretion?

%From the halo accretion rate model, we estimate the mass that will be accreted near halo center. We again choose $10^{-3}R_{vir}$, corresponding to $\sim1$ pc, to determine the accretion rates. We multiply the accretion rate by $t_{ff}(r=R_s)$, using the free-fall time at halo scale radius as the characteristic halo time scale and approximating the accreted mass in the halo center. Fig.~\ref{fig:accreted} shows the resulting masses for the Bondi-Holye and free-fall accretion rate models.

%The accreted mass remains approximately constant across redshifts, especially for the free-fall model. If the upper-limit free-fall model is true, than it is not difficult to produce massive SMBHs with $\sim10^9$ M$_\odot$ at $z\sim6$ \citep{Wu2015} as enough mass can infall near the halo center. 

%\begin{figure}
%    \centering
%    \includegraphics[width=\linewidth]{mass_accreted.png}
 %   \caption{The accreted mass at $10^{-3} R_{vir}$ for $t_{ff}$ evaluated at $R_s$. We show the accreted mass assuming $\dot{M}_{Bondi}$ and $\dot{M}_{ff}$. The accreted mass remains about constant across redshifts, and if the upper limit of free-fall accretion can be reached, enough mass to form the observed high-redshift AGN and quasars can infall to the halo center.}
 %   \label{fig:accreted}
%\end{figure}
% check dex 

\begin{figure*}[!htb]
\gridline{
\fig{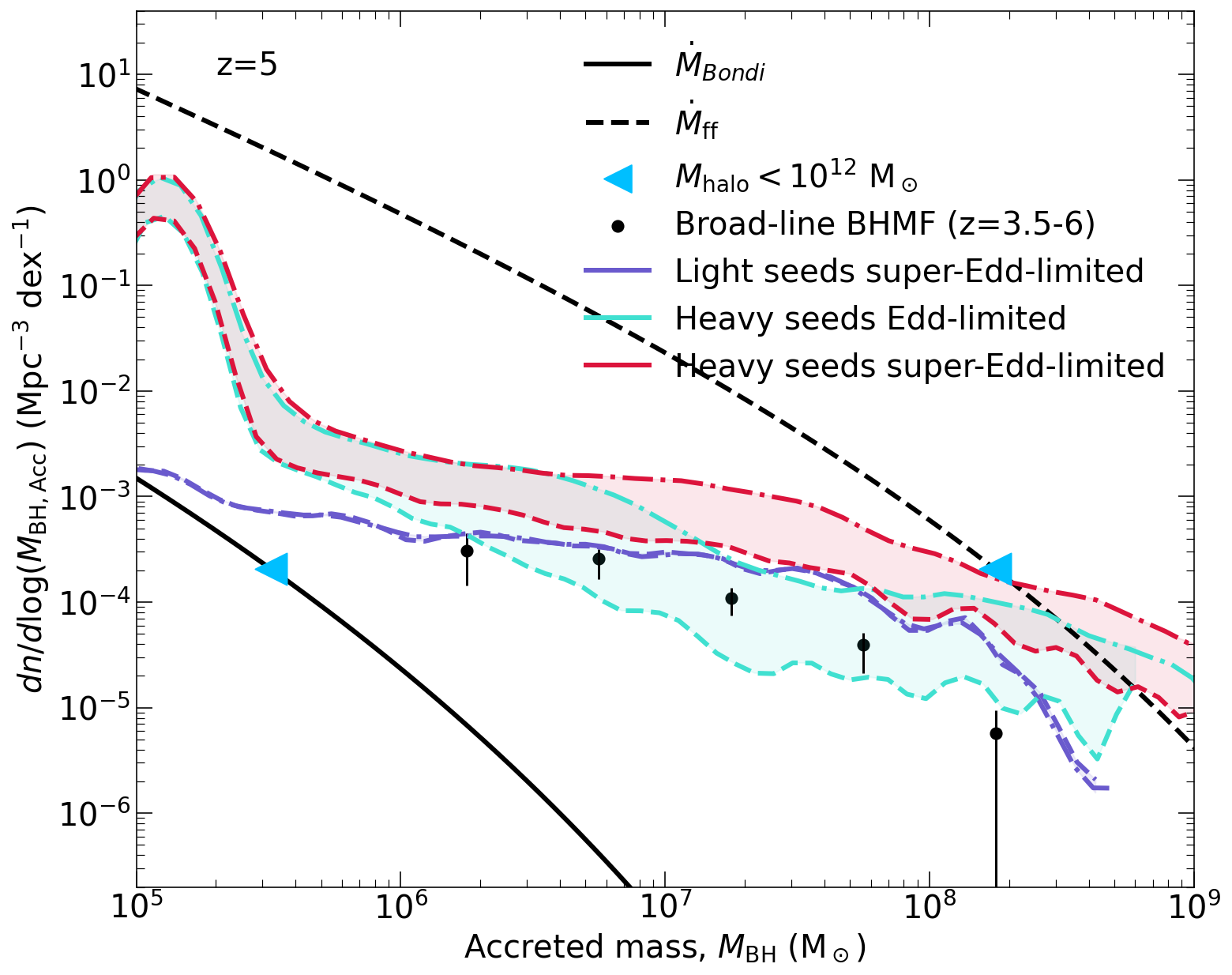}{0.5\textwidth}{}
\fig{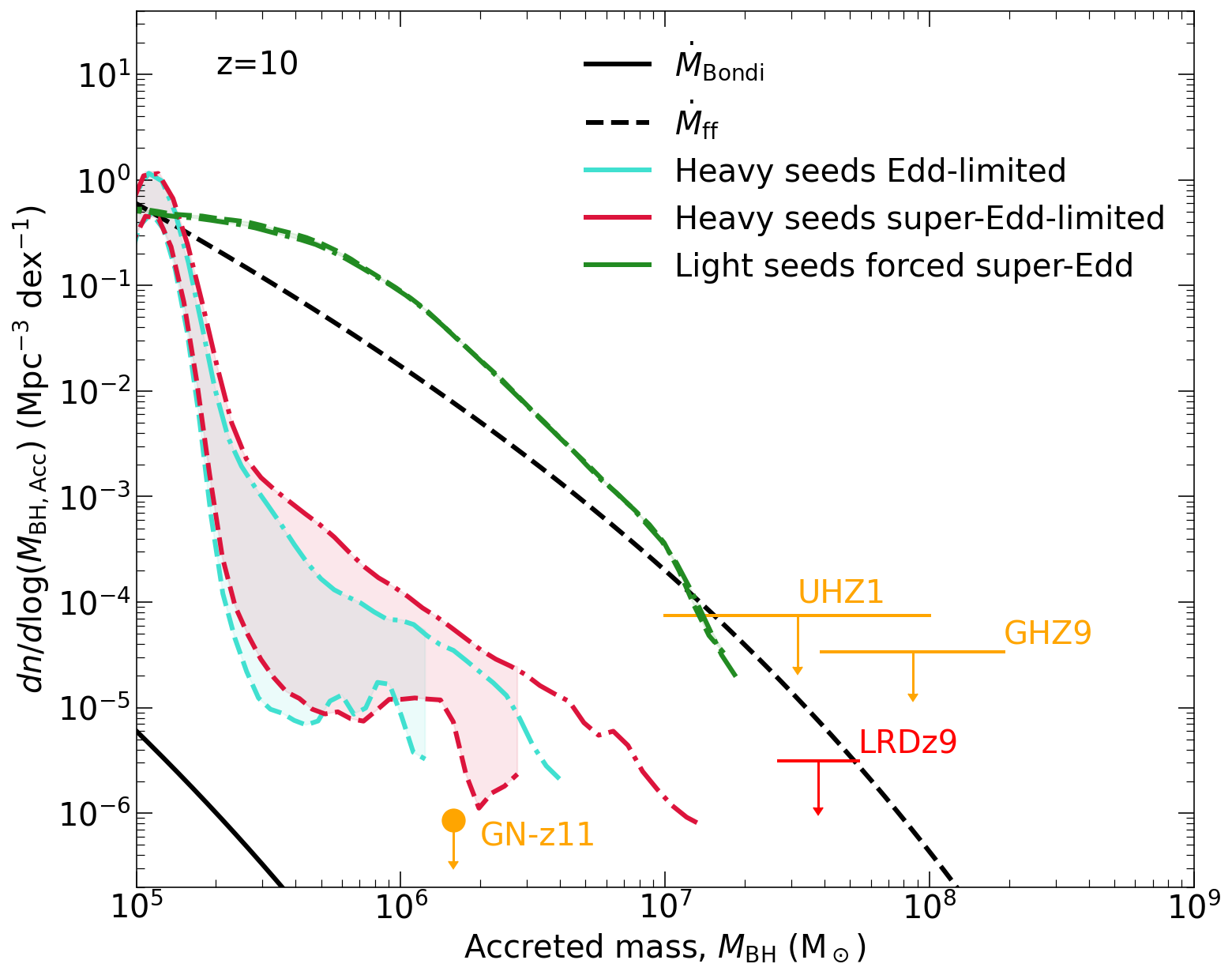}{0.5\textwidth}{}
}
\caption{Mass functions due to accretion at the Bondi-Hoyle or free-fall rates at $z=5,10$, compared with BHMFs inferred from observations \citep{Taylor2024} or theoretical modeling \citep{Jeon2025}. At $z\sim10$, the observational constraints arise from individual sources \citep{Maiolino2023,Bogdan2023,Taylor2025}, and thus are presented as upper limits. The theoretical BHMFs include models with and without heavy DCBH seeds, growing under Eddington or super-Eddington limited accretion. The distributions of accreted mass are computed similar to Fig.~\ref{fig:mvsz}, assuming that halos evolve with constant number densities. The regime of $M_{\rm halo}<10^{12}$ M$_\odot$ is marked as well. The free-fall accretion model can reproduce the most massive AGN at $z\sim10$, matching the theoretical BHMF where all SMBHs are forced to grow at super-Eddington rates (green line), but it overproduces the observed BHMF at $z\sim5$. However, the observations can only detect accreting SMBHs, and/or the accreted gas at halo centers may not necessarily be accreted by the central SMBH with AGN known to have low duty cycles \citep{Arita2025,Pizzati2025,Eilers2024}, so that the models should be considered as upper limits.}
    \label{fig:massf}
\end{figure*}

\subsection{Resulting Mass Functions}\label{sec:mf}

Given the host number densities for accreting SMBHs, as discussed above, we derive the corresponding distributions of accreted mass, and compare them with the BH mass function (BHMF) of broad-line AGN, observed at high redshifts \citep[][]{Taylor2024}, and with select $z\sim10$ AGN observations \citep{Maiolino2023,Bogdan2023,Napolitano2024,Chavez2025,Taylor2025} as shown in Fig.~\ref{fig:massf}. We also compare with theoretical BHMF predictions, for cases with and without DCBHs \citep{Jeon2025}. Whereas previously (in Figs.~\ref{fig:accnumber}, \ref{fig:mvsz}), we have considered the evolution of individual systems, the BHMF predictions assume that the entire halo population experiences gas infall at either the free-fall or Bondi-Hoyle rate.

We again find that cold-mode accretion is necessary to supply enough gas to produce the most extreme AGN. However, while the cold-mode, free fall accretion can match the approximate number densities of the $z\sim10$ AGN, at $z\sim5$, it overproduces the observed BHMF. 

As the observed BHMF represents broad-line AGN, any SMBHs that are dormant in that period will be missed. Furthermore, the infalling mass calculated in our models may not necessarily be accreted instantaneously by the central SMBH. Instead, there will likely be some delay between the gas infall to the halo center and onto the central SMBH. In fact, only around $\sim1-0.1\%$ of the free-fall mass, matching observed AGN duty cycles \citep{Arita2025,Pizzati2025,Eilers2024}, is required to match the BHMF observations. Such factors may cause the observed BHMF to exhibit abundances that are much lower than what is predicted for the free-fall model, assuming that all halos in the cold-mode regime accrete at such high rates. 

% when does the high density region disappear?
% core mass trajectory? can be expressed as constant number density instead of halo mass, which will continuously grow. How to get core mass?
% angular momentum?

\section{Lessons from High-Redshift Observations} \label{sec:discussion}

% tdes? stellar pathway?
When comparing the gas infall rates in halos to observed SMBH masses and abundances, we find that efficient cold-mode/free-fall accretion is necessary to match the masses of the observed high-redshift AGN, LRDs, and quasars. Multiple previous simulations have found similar results, that such cold inflows are necessary to fuel the central BHs and produce the observed high-redshift SMBHs \citep{Dubois2012,Dimatteo2012,Dimatteo2017,Kho2026}. A key consideration is that the masses predicted by our models are the \textit{upper limits} of SMBH growth, reflecting the amount of material that can be transported to halo centers $(10^{-3}R_{\rm vir})$. There may be a delay between depositing the gas in the halo center and the subsequent accretion onto the central SMBH, as is suggested by the low AGN efficiencies/duty cycles (see Sec.~\ref{sec:mf}). Other factors can further limit the accreted mass, in particular the effect of angular momentum hindering gas infall, or star formation within the infalling gas. Such limitations on accretion are required in any case, as a scenario where every halo experiences optimal free-fall accretion to grow its SMBH would overproduce the massive AGN (see Fig.~\ref{fig:massf}). Dark matter enhanced acceleration at high-redshifts, one condition invoked here to justify free-fall accretion even in the presence of stellar feedback, has originally been proposed to enable efficient star formation \citep{Boylan2025}. Small-scale gas physics, not realistically represented in analytical models, may determine whether efficient baryonic accretion at early times leads to SMBH feeding or star formation \citep[e.g.,][for PBH seeding]{Zhang2026}, to be further investigated in future work. However, the goal here was not to find specific high-redshift SMBH accretion pathways, but to explore the conditions for supplying enough fuel from the cosmic web to sustain efficient SMBH growth. %The masses shown in this paper thus should be considered as the optimal SMBH evolution.%, which may be achievable when the dense dark matter in high-redshift halos make stellar feedback ineffective in disrupting the gas.

Considering SMBH growth after the initial seeding is paramount, as even under heavy seed scenarios, the mass that needs to be subsequently accreted to reach the observed AGN range is orders of magnitude higher than the initial seed mass \citep{Jeon2024}.
While (super)Eddington accretion has been invoked in (semi-) analytic models to grow the BH seeds \citep{Jeon2025,Bonoli2025}, achieving such rates for extended periods is still a challenge in both cosmological and single-galaxy simulations \citep[e.g.,][]{Shin2025,Ortame2026,Mehta2026}. Thus, we discuss our models in comparison to three classes of observed SMBHs (albeit with some overlap between them): high-redshift quasars, JWST-observed AGN, and LRDs. For the most massive quasars, the rarest halos can support free-fall rates high enough to sustain their growth. However, such systems reach the fiducial halo mass limit of free fall accretion ($10^{12}$ M$_\odot$) before the quasars are observed at  $z\sim5-6$. Previous studies have found that the critical halo mass for cold mode accretion may increase with redshift \citep{Waterval2025,Keres2009}, such that the extreme quasar-hosting halos could support cold-mode accretion beyond the $10^{12}$ M$_\odot$ limit at high-redshifts.

%it is a challenge to produce them even with our upper limit accretion rates. Then, other factors may cause even more efficient gas inflow. Cold mode accretion may last to even higher halo masses than the fiducial Cold-mode accretion is shown in simulations to occur through filaments \citep{Dekel2009,Dekel2006,Birnboim2007,Keres2005}, while we have approximated spherical accretion for the free-fall accretion. If filamentary accretion can transport more massive amounts of gas than free-fall spherical accretion, the observed massive quasars could be produced. At lower redshifts, bar instability may also drive gas into the SMBH \citep{Kataria2024}. 

In contrast, JWST-observed AGN are generally less massive than quasars, and our upper limit estimates can produce them. Furthermore, many high-redshift AGN are observed to be `overmassive', where the SMBH to galaxy stellar mass ratio is much higher than in the local Universe, subject to uncertainties in mass measurement methodology \citep[e.g.,][]{Pacucci2023}. If the infalling free-fall gas is mostly accreted by the SMBH instead of forming stars, such overmassive systems could arise naturally. More detailed investigation into the nature of gas infall and star formation activity within cold inflows will be needed to confirm such behavior, however. In addition, current AGN masses at $z\sim10$ can be well matched by free fall accretion within rare $(\sim10^{-6}$ Mpc$^{-3}$) halos as shown in Fig.~\ref{fig:mvsz}. Two of them, UHZ1 and GHZ9, are found in the same survey field of the GLASS-JWST Early Release Science Program \citep{Napolitano2025}, which is consistent with predictions that massive halos should be clustered \citep{Pujol2017}.\footnote{We note that the AGN nature of UHZ1 has been brought into question \citep{Zou2026}. We include it in our discussion for completeness, but future developments may change the discussions regarding this object.} GN-z11 at $z\sim11$ is an unusual object, with a much smaller volume density than the other $z\sim10$ AGN (see Fig.~\ref{fig:massf}), but also lower in observed AGN mass. As less massive objects are expected to be more common, another explanation for GN-z11 could be that it resides in a system where efficient free-fall accretion is not possible. GN-z11 can instead originate from Bondi-Hoyle gas accretion in rare systems (Fig.~\ref{fig:mvsz}), agreeing with both its rarity and lower mass AGN. However, the $z\sim10$ AGN observations are still scarce, and future observations could challenge this picture. It has been suggested that clustering observations could identify whether the high-redshift luminous galaxies observed with JWST are the average or stochastic outliers \citep{Munoz2023}. Similarly, clustering observations in the near future could identify whether these $z\sim10$ AGN are the norm or the extreme cases. The Roman Space Telescope is expected to discover a multitude of AGN at $z\sim6-10$ \citep{Zhang2024} and will provide further constraints on theoretical predictions for possible formation pathways of high-redshift AGN. %(Hyun et al. 2026). 
Moreover, other explanations for the unique nature of GN-z11 are possible, such as PBH scenarios \citep{Maiolino2025}.
%!!! ansh paper?

Comparing our results to LRDs, if LRDs are super-Eddington accretion AGN, most of the inflowing gas will have to be accreted by the SMBH to match the accretion rates (Fig.~\ref{fig:accnumber}). This aligns with the interpretation that LRDs are AGN surrounded by dense gas with a subdominant stellar component \citep{Naidu2025,deGraff2025} as the gas will mostly be accreted by the SMBH instead of forming stars. Furthermore, the decrease in observed LRDs at low redshifts can naturally be explained by host halos becoming more massive \citep{ZhangC_2026} so that efficient cold-mode accretion is less likely and the feedback becomes disruptive as dark matter density decreases, meaning that super-Eddington accretion/dense gas cannot be sustained (see Fig.~\ref{fig:accnumber} and Sec.~\ref{sec:ratedis}). Under the fiducial choice of LRD SMBH mass, $M_{\rm BH}\sim10^6$ M$_\odot$, accreting at the Eddington rate, the size of the dense gas region is $\sim500-1000$ AU \citep{Inayoshi2025_lrd}. With hydrogen gas density of $\sim10^9$ cm$^{-3}$, the total gas mass assuming a covering fraction of unity is $\sim5$ $M_\odot$, which under Eddington accretion will be depleted in $\sim100$ years. If the SMBH is accreting at super-Eddington rates or the covering fraction is not unity, the depletion timescale will be much shorter. Thus, to sustain LRD-like characteristics, a reservoir of massive gas will be necessary to maintain the dense gas region \citep{Sneppen2026}, which the free-fall gas inflow can provide. The free-fall halo interpretation of LRDs further aligns with observed LRD masses requiring more common halos under the free-fall model (Fig.~\ref{fig:mvsz}), in agreement with their observed abundance at higher redshifts. The lack of LRDs at even higher redshifts $(z\gtrsim8)$ could be explained by the number density of halos able to support high free-fall rates for LRD-mass objects being much lower at those times.

% contradictory statements?
We also compare our results against DCBH seeding predictions, as sustained high accretion $(\sim1$ M$_\odot$ yr$^{-1}$) could naturally form DCBHs \citep{Wise2019}. However, the number densities of such highly accreting systems lie much below the predicted Eddington-accreting DCBH number (Fig.~\ref{fig:accnumber}). This could reflect that the natal structure for DCBH seeding may not follow the NFW profile, and may instead be a denser, self-gravitating gas cloud, or indicate that efficient accretion is not the only pathway to heavy seed formation in the early Universe, should they exist more abundantly. Such DCBHs may remain dormant and not grow to extreme masses after formation, unable to reach high accretion rates. We note that other works have found lower critical accretion rates ($\sim0.01$ M$_\odot$ yr$^{-1}$) for DCBH/SMS formation \citep{Kiyuna2023,Kiyuna2024}. In particular, the SEEDZ simulations \citep{Prole2026,Prole2026_2,Prole2026_3} assumed SMS formation under sustained accretion rates of $>0.02$ M$_\odot$ yr$^{-1}$, finding SMS number densities of $0.1$ cMpc$^{-3}$, higher than our prediction for the number density of systems with $\dot{M}_{\rm ff}=10^{-2}$ M$_\odot$ yr$^{-1}$ (Fig.~\ref{fig:accnumber}). Such a difference may indicate that key physics of DCBH/SMS formation occurs in a denser region than implied by the ideal NFW profile.

An interesting coincidence is that the LRD number density matches the free-fall rate $\sim1$ M$_\odot$ yr$^{-1}$ number density, which could form DCBHs (Fig.~\ref{fig:accnumber}). As DCBHs have been proposed as origins of LRDs \citep{Jeon2025_lrd,Pacucci2026,Cenci2025}, both the initial formation and the compact nature of LRDs could be explained by halo systems able to support efficient gas inflow. 
% more predictions dcbh self gravitating cloud halo mass evolution

\begin{figure}
    \centering
    \includegraphics[width=\linewidth]{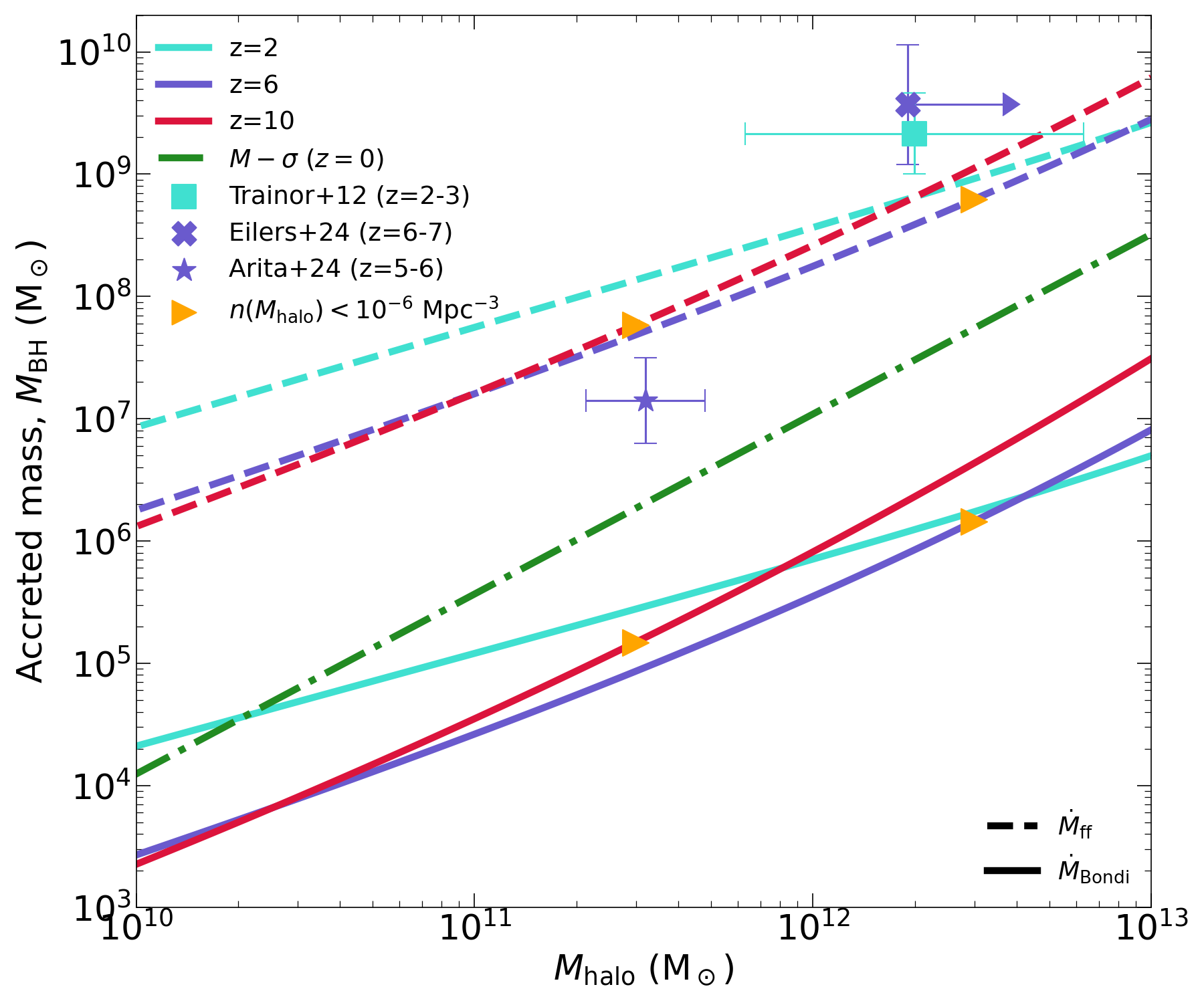}
    \caption{Accreted BH mass (approximately measured at $10^{-3}R_{\rm vir}$) vs. host halo mass for select redshifts, assuming that halos evolve with constant number densities, for the Bondi-Hoyle or free-fall rates. The free-fall rates mark the limit of the accreted mass/the maximum BH mass possible. BH mass values above the free-fall limit, the most optimal accretion, thus cannot be accounted in this way. We compare against clustering measurements of quasar \citep{Eilers2024,Trainor2012} and low-luminosity AGN \citep{Arita2025} host halo masses, and the local $M-\sigma$ relation \citep{Evrard2008,Kormendy2013}. We also mark the halo masses for which the volume density falls below $10^{-6}$ Mpc$^{-3}$ at the plotted redshifts. The high-redshift AGN host halo mass (and lower limit) measurements are below or within the free-fall boundary, but lie significantly above the Bondi-Hoyle model. Thus, to produce the observed AGN and quasars, near free-fall growth is necessary.}
    \label{fig:mhvma}
\end{figure}

There are several caveats that need to be considered here. We have assumed a spherical accretion model in this paper, while the actual accretion within halos is most likely more complex, such as accretion through filaments. Furthermore, we have not considered the role of mergers in SMBH growth, which may contribute a significant fraction to SMBH mass growth at high redshifts \citep{Bhowmick2024}. We emphasize again that our mass accretion and growth estimates are upper limits. The SMBH, should it exist at the halo center, might not accrete all the infalling material. Even under optimal accretion, AGN feedback could disrupt the cold flows \citep{Dubois2013_2}. Furthermore, we have argued that the massive and rare halos at early times match the high-redshift quasars in our analysis. The number density of such rare halos depends more strongly on the chosen cosmology and HMF, with numerical simulations showing that analytic HMFs, such as the one used here, may overestimate the number of massive halos at early times \citep[e.g.,][]{Yung2024}. When accounting for these effects, all our estimates, accretion rate number densities and mass growth, could become lower, but this would only exacerbate the problem of massive SMBHs at early times, unless extreme merger rates are invoked. 

In Fig.~\ref{fig:mhvma}, we compare the accreted mass in halos under the Bondi-Hoyle or free-fall model at individual redshifts with quasar \citep{Eilers2024,Trainor2012} and low-luminosity AGN data \citep{Arita2025}. For reference, we also reproduce host halo masses from clustering measurements and the local $M-\sigma$ relation \citep{Evrard2008,Kormendy2013}. As the free-fall model represents the most optimal accretion possible, SMBH masses above the free-fall models in a given halo mass will be forbidden, at least via gas accretion channels. The high-redshift AGN host halo masses align with the free-fall model but lie much above the Bondi-Hoyle model, indicating that free-fall growth is necessary to produce the observed SMBHs. Thus, we argue that such efficient supply of material into halo centers, and the biased halos enabling efficient accretion of gas onto the SMBH, may be necessary. However, other works that investigate SMBH accretion directly through general-relativistic magneto-hydrodynamic (GRMHD) simulations in connection to the larger cosmological environment have found that it is difficult to accrete the material onto the BH, even if the material is available \citep{Milos2009,Su2026,Cho2025}. Future work will be necessary to fully understand the mystery of efficient SMBH growth beyond the halo scales. % Jeon et al in prep?

% filaments?

\section{Summary and Conclusions} \label{sec:conclusions}
In this work, we have created analytic models to estimate gas inflow rates inside dark matter halos. We target the high-redshift environments enabling cold-mode accretion, considered the most efficient accretion possible, assuming that the baryonic gas falls in at the free-fall rate. The second model we consider is the hot mode Bondi-Hoyle accretion, following the halo temperature profile as the infalling gas is shock-heated, resulting in much smaller aacretion rates. Combining these models with the cosmological HMF across redshifts, we examine the accretion rate distributions across redshift (Sec.~\ref{sec:ratedis}), the mass evolution following individual halos (Sec.~\ref{sec:accmass}), and the mass functions of the accreted BH masses (Sec.~\ref{sec:mf}). 

In comparisons to observations, we find that most observed high-redshift SMBHs could be explained by the cold-mode free-fall accretion that can naturally occur in available high-redshift, less-massive halos, such that the efficient gas inflow is supplying enough fuel to be able to grow to the large inferred masses. Overmassive systems could arise from the cold inflowing gas being accreted by the SMBH instead of forming stars. The most massive quasars require the rarest and most massive halos, as expected. Less massive high-redshift AGN observed by JWST require less massive and more common halos, although under the Bondi-Hoyle model their number densities will be much lower than observed. For LRDs, we find that their number densities align with halos able to support super-Eddington accretion/dense gas regions under free fall, in agreement with models that explain LRDs through super-Eddington accreting and/or enshrouded AGN \citep{Madau2026,LiuH2025,Secunda2026,Pacucci2024,Inayoshi2025_bl,Kokorev2025}. LRD's low number density at low redshifts could be explained by the cold-mode accretion shutting off as the halo mass grows and feedback becomes more efficient. At higher redshifts, the halos able to host LRD-mass objects are also rare so that the LRD number density becomes lower as well. 

In comparison with theoretical DCBH number density predictions \citep{Jeon2025}, if DCBHs form through sustained accretion, we find a much lower number density than the predicted upper limits. This may indicate an actual dearth of DCBH formation, multiple formation pathways for heavy seeds, or that DCBH/SMS formation occurs on smaller and denser scales than the larger halo scales considered in this work \citep{Kiyuna2024,Prole2026}. 

Follow-up theoretical work will be needed to test the models in numerical simulations and on smaller scales. The analytic estimation presented here most likely will not apply at small-scales with nonlinear dynamics when approaching the SMBH event horizon. However, to resolve the issue of both the gas inflow and SMBH accretion, large-volume and high-resolution simulations will be needed to simulate gas inflow and SMBH accretion. Other works have started with resolving the actual SMBH accretion and bridging the scales to the larger cosmological environment \citep{Su2026,Cho2025}, but found that resolved SMBH accretion is lower than the Bondi-Hoyle approximation. Determining which systems can host efficiently accreting SMBHs will be key to understanding the high-redshift SMBHs. 

Observationally, JWST and Roman will find more AGN at high redshifts. High-redshift $(z\sim10)$ AGN and/or lower mass AGN observations will provide more insight to the formation of the first SMBHs \citep{Jeon2025,Fei2025}. In turn, the required evolution pathways for such SMBHs to become the massive quasars as observed will be more constrained. As more and more massive AGN are found at high-redshifts, we may need to consider them not as extreme systems arising from random distributions of biased systems, but as a natural consequence of cosmological structure formation. The efficient gas inflow to dark matter halos at high-redshift predicted for star formation could also feed the BHs, providing natural pathways for massive SMBHs, overmassive systems with subdominant stellar masses, and LRDs.

% scatter in mhal vs mdot. When does transition happen to bondi. simplify the plots since bondi is minimal? add model plot?

% lack of lrd at high z


\section*{Acknowledgments}
VB acknowledges support from the Josey Centennial Professorship in Astronomy at UT Austin. MBK acknowledges support from NSF grant AST-2408247; HST-GO-16686, HST-AR-17028, JWST-GO-03788, and JWST-AR-06278 from the Space Telescope Science Institute, which is operated by AURA, Inc., under NASA contract NAS5-26555; and from the Samuel T. and Fern Yanagisawa Regents Professorship in Astronomy at UT Austin. JBM acknowledges support from NSF Grants AST-2307354 and AST-2408637, and NASA through grant JWST-GO-03224.

%\vspace{5mm}


%\software{}




\bibliography{ms}{}
\bibliographystyle{aasjournal}



\end{document}